\documentclass[preprint,review,12pt]{elsarticle}

\usepackage{amssymb}

\usepackage{lineno}
\usepackage{siunitx}
\usepackage{multirow}
\usepackage{amsmath}
\usepackage[capitalize]{cleveref}
\usepackage{xcolor}
\usepackage[section]{placeins}
\usepackage{float}
\usepackage{nomencl}
\makenomenclature
\usepackage{adjustbox}
\usepackage{longtable}

\journal{Combustion and Flame}

\begin{document}
	
\begin{frontmatter}



\title{Thermal inertia effect of reactive sources on one-dimensional discrete combustion wave propagation}


\author{Daoguan Ning\corref{cor1}}
\ead{d.ning@tue.nl}

\author{Yuriy Shoshin}

\address{Department of Mechanical Engineering, Eindhoven University of Technology, the Netherlands}

\begin{abstract}
\if
The discrete flame is an unique combustion phenomenon encountered in heterogeneous systems consisting of usually a continuum gaseous oxidizer and condensed point-like reactive sources (e.g., fuel particles). Since the density of the reactive sources is commonly three orders of magnitude larger than that of the gas, they have considerable thermal inertia. In spite of the thermal inertia, the propagation properties of the the discrete flame, such as flame speed and propagation limits, have been predicted by a discrete thermal equilibrium model (DTEM) with the assumption of the thermal equilibrium between the gas and the condensed reactive sources. 
\fi

\textcolor{black}{In the present work, the discrete flame model \cite{goroshin1998effect} is augmented by introducing the thermal inertia of particles in the preheating zone.}
The effect of particle thermal inertia on flame speed, propagation limits, and near-limits dynamics of one-dimensional discrete combustion waves is studied using the new model. It is found that, with the increase of particle thermal inertia, the propagation velocity of the discrete flame decreases due to a smaller heating rate of the particles. Besides, particle thermal inertia extends the propagation limits compared to the prediction of the old model. Furthermore, it is mathematically proven that the nonphysical branch of the solutions for the discrete flame speeds, found using the old discrete model, is a set of solutions for the propagation limits of steady-state discrete flames with particle thermal inertia included. The flame speed predicted using the new model is also compared with that determined analytically using a \textcolor{black}{continuum} model considering the thermal inertia of the condensed phase \cite{goroshin1996quenching}. We find that the discrete flame speeds predicted by the both models become closer to each other with increasing particle thermal inertia. Finally, the two models converge regardless of the discrete nature of the heat sources when particle thermal inertia is large enough so that can limit the flame propagation. The particle thermal inertia controlled flames could be regarded as a new kind of combustion regime.
\end{abstract}





\end{frontmatter}
\section*{\textcolor{black}{Nomenclature}}
\textcolor{black}{
\begin{longtable}[l]{lll}
	$A$ && particle surface area [\SI{}{m^2}]\\
	$B$ && particle concentration [\SI{}{kg/m^3}]\\
	$c$ && specific heat  [\SI{}{J/kg/K}] \\
	$h$ && heat transfer coefficient [\SI{}{W/m^2/K}]\\
	$\textbf{\emph{H}}$ && Heaviside function \\
	$i$ && index of particle \\
	$l$  && mean inter-particle distance [\SI{}{m}] \\
	$N$ && number of particles \\
	$r$ && particle radius [\SI{}{m}] \\
	$t$ & & Dimensional time [s]   \\  
	$V$ && particle volume [\SI{}{m^3}]\\
	$\bold x$ && dimensional coordinate vector \\
	$y$ && dimensionless spacial coordinate \\    
\end{longtable} }

\subsection*{\textcolor{black}{Greek symbols}}
\textcolor{black}{
\begin{longtable}[l]{lll}
	$\alpha$ && thermal diffusivity [\SI{}{m^2/s}] \\
	$\gamma$ && dimensionless particle thermal inertia, defined \cref{eqn: discrete particle equation} \\ 
	$\Delta$ && time interval \\
	$\kappa$ && squared dimensionless flame speed \\
	$\lambda$ && thermal conductivity [\SI{}{W/m/K}] \\
	$\theta$ && dimensionless temperature \\
	$\rho$ && density [\SI{}{kg/m^3}] \\
	$\tau$ & & Dimensionless time \\
	$\xi$   && defined in \cref{non-dimentinal governing euqation for particle 1}, $\xi=\rho_{s} c_{s} r^2_p/(3\lambda_{g,u} t_c)$ \\
	$\omega$ && dimensionless heat source term \\	
\end{longtable} }

\subsection*{\textcolor{black}{Subscripts}}
\textcolor{black}{
\begin{longtable}[l]{lll}
		$a$ && adiabatic   \\
		$c$ & & combustion \\ 
		$d$ && heat diffusion in the gas phase between particles \\  
		$e$ && heat exchange between gas phase and particles \\     
		$g$ && gas \\  
		$ign$ && ignition \\ 
		$p$ && particle \\
		$u$ && unburned mixture at $y \to -\infty$ \\
\end{longtable} }


\section{Introduction}
\label{sec 1}
Flame propagation in heterogeneous media consisting of a continuum gas phase and condensed reactive sources is an important phenomenon. It is encountered in a variety of practical processes, such as burning of liquid fuel spray and combustion of solid particle aerosols including coal, biomass, and metals. Depending on the heterogeneous system, the size of the reactive source varies from mircon-sized particles to millimeter-sized droplets. For low-volatile condensed fuels, exothermic chemical reactions happen at or close to the interfaces between localized reactive sources  and a continuum gaseous oxidizer, which leads to spatially discrete heat sources. In spite of the discrete nature of the heat sources, theoretical models developed before early 90s, as typical examples \cite{rumanov1971flame,krazinski1979coal,ballal1983flame}, commonly adopted the homogeneous continuum approximation, in which the heat sources were modeled by a continuous function of spatial coordinates. These models fail to reveal important phenomena related to the discreteness of the heat sources, in particular independence of flame speed on particle combustion time for fast burning particles. 
On the contrary, continuous models, developed for both gaseous and heterogeneous systems, predict the flame speed to be proportional to the square root of the reaction rate, which is inversely proportional to the particle combustion time in the later case.

A novel theoretical model, elucidating the effects of spatial discreteness of reactive sources on the combustion wave propagation,  
was first proposed by Shoshin et. al \cite{shoshin1987flame} and further explored by Goroshin et. al \cite{goroshin1998effect, tang2009effect} and Beck and Volpert \cite{beck2003nonlinear}. The theory (i.e., discrete flame model) idealizes reacting particles as point-like heat sources based on the fact that the mean distance between the particles is around two orders of magnitude larger than their sizes in fuel lean suspensions. 
Besides, the model introduces a stepwise kinetic, i.e., the exothermic reactions are triggered when a fixed ignition temperature is reached. After ignition, the particles combust at a constant reaction rate within a burn time.
The stepwise kinetic is a simplified physical representation of reactive particles burning in the external diffusion regime \cite{soo2018combustion}.
The temperature field of the continuum phase is then obtained through superimposing heat diffusion waves, generated by every point-like heat source, using the Green's function and  the exact spatial coordinate of each heat source. Thus, the continuum approximation is eliminated. An analytical solution for the flame propagation velocity 
can be derived from the model when mono-dispersed particles are regularly distributed in space. 
For systems with randomly located point-like reactive particles, the mean propagation speed and the front structure of a combustion wave can be obtained by computer-assisted numerical simulations for systems with adiabatic \cite{goroshin2011reaction, tang2012propagation, lam2017front} or conductive boundaries \cite{lam2020dimensional}.

The discrete flame model provides a criterion to differentiate between the discrete and \textcolor{black}{continuum} combustion waves. It points out that owing to the spatially discrete nature of the heat sources, a heterogeneous flame can propagate in \textcolor{black}{continuum} or discrete regimes
depending on the dimensionless combustion time (also known as the discreteness parameter \cite{goroshin1998effect}):

	\begin{equation}
 		\tau_c = \frac{t_c}{t_d} = \frac{ t_c\alpha_u}{l^2},
	\end{equation}
which is the ratio between the physical combustion time of an individual reactive particle $t_c$ (also the inverse of chemical reaction rate) and the average inter-particle heat diffusion time $t_d = l^2/\alpha_u$ ($l$ is the average distance between neighboring particles and $\alpha_u$ is the mean thermal diffusivity of the fresh gas-particle mixture). When $\tau_c \gg 1$, the chemical reaction is the slowest process that limits the combustion wave propagation. Heat generated by individual sources has enough time to dissipate among them, and the flame exhibits the features of a continuum combustion wave. Therefore, continuum theories such as \cite{rumanov1971flame,krazinski1979coal,ballal1983flame},  developed based on the approximation that the control volume still contains a large number of sources, are valid. 
When $\tau_c \ll 1$, the chemical reaction rate is much faster than the heat diffusion rate. The flame propagation is controlled by the inter-particle heat diffusion instead of the chemical reaction. In contrast to a \textcolor{black}{continuum} combustion wave, this new kind of flame is called a discrete combustion wave. A lot of experimental evidence has been found to prove the existence of discrete combustion waves, which is partly summarized in a review paper \cite{mukasyan2008discrete}.

The propagation limit of a discrete combustion wave also differs from that of a \textcolor{black}{continuum} flame. For a \textcolor{black}{continuum} combustion wave, its propagation limit is constrained by the thermodynamic limit (where ignition temperature equals adiabatic flame temperature) in the absence of external heat losses. In contrast, analysis conducted by Tang et al. \cite{tang2009effect}, using the discrete model for a regular array of particles, shows that the propagation limit of a presumed steady-state discrete combustion wave is considerably below the thermal dynamic limit. The nature of this propagation limit will be discussed in the following sections. Furthermore, as the theoretical propagation limit is approached, the flame exhibits complex dynamics and can quench due to the chaotic flame speed. Rashkovskiy et al. \cite{rashkovskiy2010one} performed detailed transient numerical simulations based on the same model for the dynamic behavior of  one-dimensional discrete flames. Their results show that the flame behavior transits from stable propagation to periodical bifurcation and further to chaotic fluctuations with increasing ignition temperature.

\if
As the first approximation is only reasonable when a ignition  temperature exists, the applicability of the solution is limited to the sources burning in the diffusion regime. However, there is a significant temperature difference between the gas and the condensed phase when particles burn in the diffusion regime. In the preheating zone, the particles behaves as a heat reservoir. In contrast, they are heat sources in the combustion zone.  At first glance, the second approximation seems incorrect. However, similar to the discreteness parameter, 
\fi

The discrete model incorporates the simplification of thermal equilibrium between the gas phase and condensed particles in the flame preheating zone, which is based on the assumption that the heat transfer rate between them is much faster than other physical processes in the flame front. Therefore, the discrete model developed in \cite{shoshin1987flame} will be referred to as discrete thermal equilibrium mode (DTEM).
The rate of reaching the thermal equilibrium (interphase heat transfer rate) is inversely proportional to the particle thermal inertia (cf. Appendix A), which represents the characteristic time scale of the heat exchange between the gas and a particle. 
An estimation of the value of particle thermal inertia for common metal dust suspensions, such as Al and Fe, shows that the interphase heat transfer rate is in the same order of magnitude as inter-particle heat diffusion rate. Therefore, the approximation of thermal equilibrium could be unjustified for some cases. The \textcolor{black}{continuum} model with thermal equilibrium assumption (CTEM) reported in \cite{goroshin1998effect}, used to compare with DTEM for the identification of the discreteness effect,  has been further extended by including the particle thermal inertia in the preheating zone \cite{goroshin1996quenching, goroshin2000flame}, which yields a \textcolor{black}{continuum} thermal inertia model (CTIM). However, the extension has not been implemented into DTEM until the current work.

\if
This assumption is only applicable when the heat transfer rate between the two phases is far faster than all other processes 

The theoretical model developed for the propagation velocity based on the homogeneous reaction source assumption is only valid when $t_c \gg t_d$

 One of the key future of flames that differentiates the 
 
 Experimental and theoretical advances made in last two decades have brought a new kind flame into horizon, so-called discrete combustion wave. 
 The condensed phase and the continuum phase 
 
 At the same time, the reacting cells also serve as discrete heat reservoirs before ignition.
 \fi
 
In this work, we will extend the discrete flame model reported in \cite{goroshin1998effect,shoshin1987flame} by introducing an extra heat balance equation for the condensed reactive particles, where the value of particle thermal inertia is included as an adjustable parameter.
The extended discrete mode will be referred to as discrete thermal inertia model (DTIM).
As in \cite{shoshin1987flame, tang2009effect, beck2003nonlinear, rashkovskiy2010one}, a one-dimensional approximation for a combustion wave propagating in a system of regularly distributed mono-dispersed reacting particles will be considered. The effect of the discrete nature of reactive particles on flame speed will be re-examined after considering the particle thermal inertia. To achieve this goal, we will compare the flame speed obtained from DTIM and CTIM. Furthermore, the thermal inertia effect on the propagation dynamics and limits of discrete combustion waves for regularly distributed particles will be investigated systematically and compared with earlier results for the zero-thermal-inertia case. 


\section{Mathematical models and numerical treatment}

\textcolor{black}{In order to provide a background knowledge to the readers of the present study, CTEM, CTIM, and DTEM will be concisely re-introduced in \cref{subsec: continous model,subsec: discrete model}, respectively. The detailed model description and derivation of associated equations can be found in \cite{goroshin1998effect,goroshin1996quenching,goroshin2000flame}, and therefore not repeated here. The introduction of particle thermal inertia into the discrete model is then presented.}

\subsection{Continuum model and steady-state solution}
\label{subsec: continous model}
 If the widths of both the flame reaction and preheating zones are sufficiently larger than the average distance between neighboring particles, i.e. $\tau_c\gg1$, the smallest control volume adopted by the continuous model still contains a large number of particles  \cite{goroshin1998effect}. Therefore, the heat source can be modeled based on the continuum approximation using average heat release rate of particles in the reaction zone multiplied by the fuel mass concentration.
\textcolor{black}{Furthermore, the assumptions are adopted as follows:}
\begin{enumerate}
    \item \textcolor{black}{The oxidizer in the mixture is abundant.}
	\item \textcolor{black}{The heat capacity of the mixture is constant.}
	\item \textcolor{black}{Particles and gas are at the thermal equilibrium state.}
    \item \textcolor{black}{The thermal conductivity of the mixture linearly dependents on temperature.}
   
\end{enumerate}
The dimensionless linear equation and the boundary conditions governing a steady-state \textcolor{black}{1D} flame propagation can be expressed as \cite{goroshin1998effect,seshadri1992structure}:

\begin{equation}
	\label{non-dimentinal governing euqation}
	\frac{d^2 \theta_g}{dy^2}=\kappa\frac{d\theta_g}{dy}+\omega,
\end{equation}
\begin{center}
	$y \to -\infty:\ \theta_g=0,\ $
	$y=0:\ \theta_g = \theta_{g,ign},\ $ 
	$y=1:\ d\theta_g/dy=0,$
\end{center}
where  $\theta_g = \frac{T_g-T_u}{T_a-T_u}$ is the dimensionless gas temperature;  $T_g$, $T_u$, and $T_a$ are the dimensional temperatures of the gas, the fresh mixture, and the adiabatic flame, respectively; $y = x/(v_ut_c)$ is the dimensionless spacial coordinate, where $x$, $v_u$, and $t_c$ are the streamwise coordinate, flame speed and particle combustion time, respectively; $\kappa = v_u^2t_c/\alpha_u$ is the squared dimensionless flame speed, and $\alpha_u$ is the thermal diffusivity of  the unburned upstream mixture; $\omega$ is the dimensionless heat source term, which is 0 in both the preheating zone ($y<0$) and the post flame zone ($y>1$) and equals $-\kappa$ in the combustion zone ($0 \leqslant y \leqslant 1$); $\theta_{g,ign}$ is the dimensionless gas temperature at which particle ignition occurs in a propagating combustion wave. \textcolor{black}{In fact, \Cref{non-dimentinal governing euqation} can also be derived when constant $\rho_g$ and $\lambda_g$ are assumed. This is because the thermal expansion of the gas (i.e., $\rho_g \propto T^{-1}$) compensates the change of thermal conductivity caused by temperature ( i.e., assumption 4, $\lambda_g \propto T^1$) \cite{seshadri1992structure}.}

\Cref{non-dimentinal governing euqation} can be solved separately in the preheating and combustion zones using the corresponding boundary conditions. By matching the heat flux at the boundary between the preheating and reaction zones, 
a well-known implicit solution for the dimensionless flame speed as a function of the dimensionless gas ignition temperature can be expressed as \cite{goroshin1998effect, williams1985combustion}:

\begin{equation}
	\label{eq:continous solution with gas ignition}
	\theta_{g,ign} = \frac{1 - e^{-\kappa}}{\kappa},
\end{equation}

When neglecting the particle thermal inertia, the gas ignition temperature is equivalently assumed to be the same as the particle ignition temperature, i.e., $\theta_{g,ign} = \theta_{p,ign}$. In practice, the particle temperature is always lower than the gas temperature in the preheating zone of a heterogeneous flame due to particle thermal inertia. Moreover, the particle ignition temperature is an explicit input parameter that is justified in some cases, like for Al particles, which ignite when the oxide shell melts. Whereas, the gas ignition temperature cannot be determined beforehand because it depends on the particle residence time in the preheating zone, i.e., on the flame speed.
\if at ignition event, particle temperature is known as its' ignition temperature, while gas temperature is usually not. \fi Assuming Nusselt number $Nu = 2$, the dimensionless governing equation describing heat transfer from the gas to an unburned condensed phase (multi particles in the control volume) and the boundary conditions are written as \cite{goroshin2000flame}:


\begin{equation}
	\label{non-dimentinal governing euqation for particle 1}
	\frac{d\theta_p}{dy} = \frac{\theta_g - \theta_p}{\xi},
\end{equation}
\begin{center}
	$y \to -\infty:\ \theta_p=0,\ $
	$y=0:\ \theta_p = \theta_{p,ign},\ $ 
\end{center}
where $\xi = \rho_{s} c_{s} r^2_p/(3\lambda_{g,u} t_c)$; $\rho_{s}$, $c_{s}$, and $r_p$ are the density, specific heat, and radius of an unburned particle, respectively;  $\lambda_{g,u}$ is the thermal conductivity of the fresh gas. 
An analytical expression between the ignition temperatures of gas and a particle can be found from the solution of \cref{non-dimentinal governing euqation for particle 1}:

\begin{equation}
	\label{eqn: relation between gas and particle ignition temperature}
	\frac{\theta_{g,ign}}{\theta_{p,ign}} = 1+\kappa\xi.
\end{equation}
Note that \cref{eqn: relation between gas and particle ignition temperature} slightly differs from the corrected expression\footnote{Equation 4 in \cite{goroshin2000flame} is wrong, whose left-hand side should be inverted.} reported in \cite{goroshin2000flame} as here we use a different dimensionless form for temperature. 
Finally, by substituting \cref{eqn: relation between gas and particle ignition temperature} into \cref{eq:continous solution with gas ignition}, the implicit solution for the dimensionless flame speed as a function of the dimensionless particle ignition temperature is found as:

\begin{equation}
	\label{eq:continous solution with particle ignition}
	\theta_{p,ign} = \frac{1 - e^{-\kappa}}{\kappa(1+\kappa\xi)}.
\end{equation}

\subsection{Discrete model and transient semi-numerical simulation}
\label{subsec: discrete model}
In a thermal diffusion system where point-like heat sources (e.g., fuel particles) are distributed in a continuum phase (usually a gas or a gaseous mixture),  \textcolor{black}{the heat waves generated by individual particles can be described the by Green's functions. It is assumed that the thermal diffusivity of the system is constant and the particle is defined by the $\delta$-function in the spatial coordinates. Owing to the linearity of the heat diffusion equation for the system,} an analytical expression for the dimensionless temperature filed of the continuum phase $\theta(\bold x,\tau)$ as a function of the dimensionless coordinate vector $\bold x(\frac{x}{l},\frac{y}{l},\frac{z}{l})$ and the dimensionless time $\tau = t/t_d$ can be found by the superposition of Green's functions, provided that ignition moments of burning or already burned particles are known \cite{goroshin1998effect,shoshin1987flame}:

\begin{equation}
\theta_g(\bold x, \tau)=
\begin{cases}
\displaystyle \sum_{i=1}^{N}\left( \frac{1}{4\pi\Delta\tau_i}\right) ^{p/2} \exp(-\frac{\left| \bold x - \bold x_i\right| ^2}{4\Delta\tau_i}), & \text{if}\ \tau_c=0 \\

\displaystyle \sum_{i=1}^{N} \frac{1}{\tau_c}\int_{(\Delta \tau_i-\tau_c)\textbf{\emph{H}}(\Delta \tau_i-\tau_c)}^{\Delta \tau_i} \left( \frac{1}{4\pi\tau}\right) ^{p/2} \exp(-\frac{\left| \bold x - \bold x_i\right| ^2}{4\tau})d\tau, & \text{if}\ \tau_c>0
\end{cases}
\label{eqn: discrete governing equation of gas}
\end{equation}
where N is the total number of  already ignited particles and $\Delta \tau_i = \tau-\tau_i$ is the time elapsed since the ignition of the $i$-th particle; the exponent $p=1$, 2, and 3 is for one, two, and three dimensional systerms, respectively;   $\textbf{\emph{H}}$ is the Heaviside function and $\bold x_i$ is the dimensionless coordinate vector of the $i$-th ignited particle.

An analytical solution for the dimensionless flame speed $\eta$ can be derived from \cref{eqn: discrete governing equation of gas}  with point-like heat sources distributed in a three-dimensional regular lattice by assuming a steady-state flame front  propagating from one plane of particles to the next with a constant time delay $\delta_{\tau} = 1/\eta$, where $\eta = v_ul/\alpha_u$, between the ignition events of two consecutive planes.  The detailed expression is available in \cite{shoshin1987flame, goroshin1998effect}, and is not repeated here for simplicity. It was also shown in \cite{shoshin1987flame, tang2009effect} that the above mentioned three-dimensional system can be closely approximated by a one-dimensional system of evenly spaced infinitely thin planes in which heat release occurs. In such a model, discreteness of heat sources only in the direction of flame propagation is accounted for. The reacting planes in the following will also be called ``particles". 




When neglecting the particle thermal inertia, the temperatures of a particle and the corresponding local gas are identical. Therefore, the ignition temperatures of a particle and the local gas temperature are equal, i.e., $\theta_{g,ign} = \theta_{p,ign}$. When the particle thermal inertia is considered, the ignition temperature of gas is larger than that of a particle and unknown a priori for a given particle ignition temperature $\theta_{p,ign}$.
The \cref{eqn: discrete governing equation of gas} has to be solved numerically, coupled with the governing equation for the dimensionless particle temperature described as:

\begin{equation}	
	\frac{d\theta_{p}}{d\tau}=\frac{\theta_{g}-\theta_{p}}{\gamma}.
	\label{eqn: dimensionless discrete particle temperature equation 1}
\end{equation}
where $\gamma = \frac{c_s\rho_{s}r_{p}^2}{3\overline{c\rho}l^2}$ is a newly defined dimensionless particle thermal inertia; $\overline{c\rho}$ is the average volumetric heat capacity of the system. Physically, $\gamma$ represents the ratio between the characteristic time scales of gas-particles heat exchange ($t_e$) and heat diffusion between neighboring particles in the system ($t_d$). A further detailed derivation of \cref{eqn: dimensionless discrete particle temperature equation 1} and the physical representation of $\gamma$ can be found in Appendix A. In the numerical simulation, particles ignite when their temperature reaches $\theta_{p,ign}$ for the first time \footnote{The analytical solution of DTEM suggests that the ignition temperature can be reached by a particle twice, and therefore, the particle can ignite at either time. But only the first time represents a physical solution \cite{tang2009effect}.}. 
The time of every ignition event is registered, and the local flame propagation speed can be calculated as $\eta_i = 1/\Delta \tau_{ign}$ for one-dimensional simulation, where $\Delta \tau_{ign}= \tau_{ign,i}-\tau_{ign,i-1}$ is the time delay between the ignition moments of two consecutive particles. 

The key dimensionless parameters used in the continuous and discrete models are connected via: $\kappa=\eta^{2}\tau_{c}$ and $\xi=\gamma\slash\tau_{c}$. Apparently, CTIM and DTIM collapse to CTEM and DTEM when $\gamma=0$, respectively. All physical parameters discussed in the following sections are dimensionless unless otherwise stated.
  
\section{Results and discussion}
\subsection{Propagation velocity of stable discrete combustion wave }
In this section we will first validate the numerical simulation by comparing the numerical and analytical values of the propagation velocities of one-dimensional stable discrete combustion waves without considering the thermal inertia effect. The effect of thermal inertia during particle preheating on the discrete flame propagation speed will be then examined.

\begin{figure}[h]
	\centering\includegraphics[width=10cm]{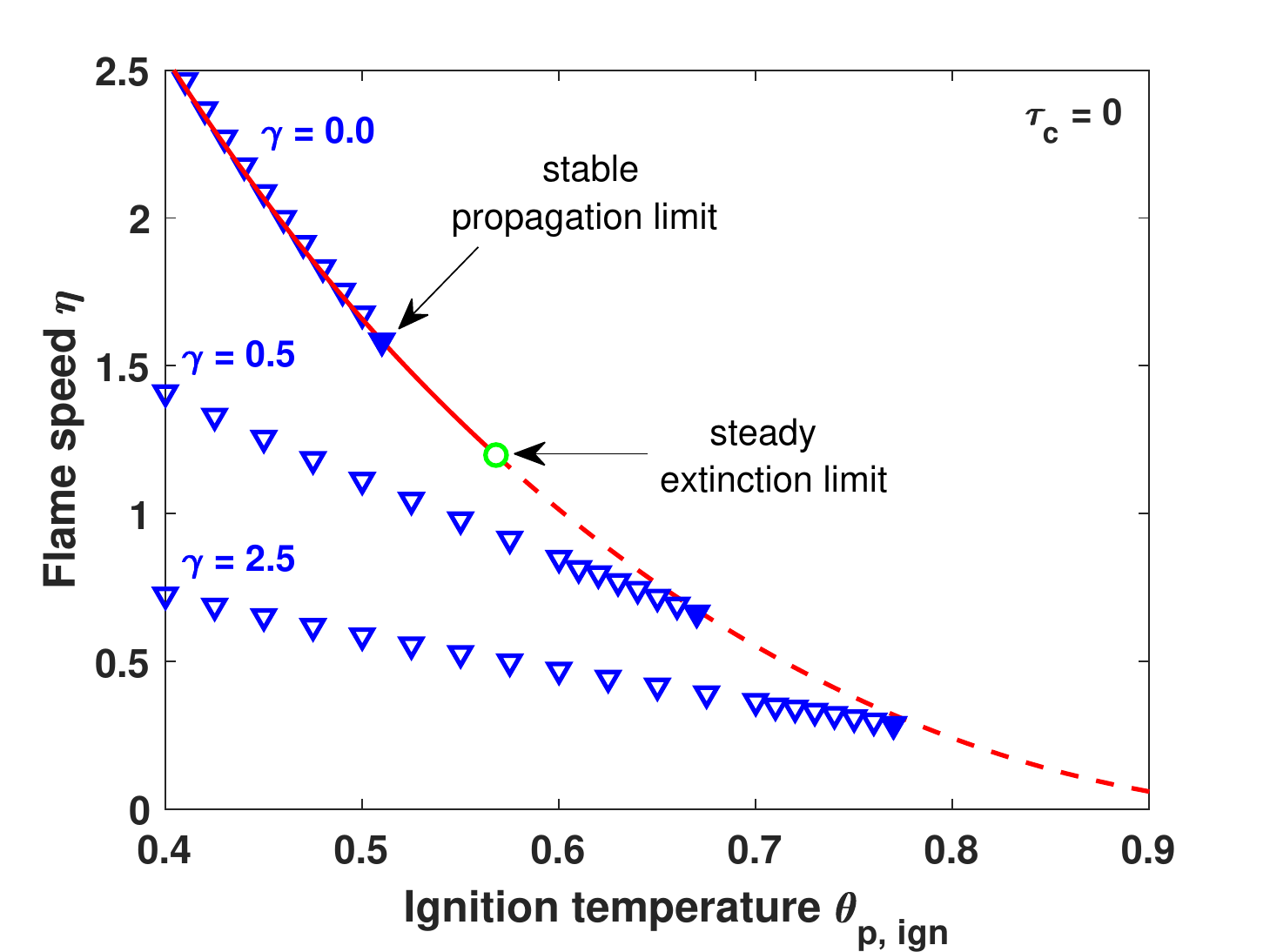}
	\caption{Flame speed as a function of particle ignition temperature; blue triangles represent transient simulation results, and the dynamical propagation limits at $\gamma=0.0,\ 0.5,\ 2.5$ for $\tau_c=0$ are indicated by solid triangles; red solid and dashed lines are physical and nonphysical solutions of the steady-state discrete model at $\gamma=0 $ for $\tau_c=0$, respectively; the green open circle is the steady extinction limit.}
	\label{fig:flame speed at tau_c of 0}
\end{figure}

For instantly burning particles without considering the particle thermal inertia during  preheating, i.e. $\tau_c=0$ and $\gamma=0$, CTEM predicts an infinite flame propagation speed owning to neglecting the inter-particle heat diffusion. In contrast, DTEM is still able to provide analytical solutions for flame propagation speeds. \Cref{fig:flame speed at tau_c of 0} shows the flame propagation speeds found analytically when $\gamma = 0$ as well as numerically computed results at $\gamma = 0$, 0.5, and 2.5 using DTIM. At $\gamma = 0$, the analytical solutions for $\eta$ as a function of $\theta_{p,ign}$ are obtained by DTEM with a steady-state assumption. However, beyond a critical ignition temperature $\theta_{ex}\approx 0.568$, a particle ignites when reaching the ignition temperature at the second time during cooling \cite{tang2009effect}. Therefore, the solutions found for $\theta_{p,ign}>\theta_{ex}$ are nonphysical. The critical ignition temperature $\theta_{ex}$,  is defined as the steady extinction limit for DTEM. The simulated flame speed $\eta$, obtained using DTIM, as a function of  particle ignition temperature $\theta_{p,ign}$ for various particle thermal inertia $\gamma$ is shown as blue markers in \cref{fig:flame speed at tau_c of 0}. At $\gamma=0$, the simulated results agree well with the analytical values for stably propagating flames. Here, stable flames are defined as flames with a speed fluctuation amplitude smaller than 1\% of its mean value. In the transient numerical simulation, the flame can still propagate but becomes unstable when further increasing the ignition temperature above the stable propagation limit until the dynamic extinction limit is reached, where the flame extinguishes due to chaotic flame speed or unbounded growth of the flame speed oscillations \cite{tang2009effect}. 
In the unstable regime, complex flame dynamics is exhibited including periodical oscillations and chaotic propagation \cite{tang2009effect}. The effect of particle thermal inertia on flame dynamics will be discussed in \cref{subsec 3.3}. Obliviously, the flame speed becomes smaller with increasing particle thermal inertia because of a slower heating rate of an unburned  particle. It is very interesting to find that the stable propagation limits extend with increasing particle thermal inertia and, upon increasing, approach the nonphysical analytical solutions for the flame speed solved with DTEM, where no particle thermal inertia is considered    ($\gamma=0$).

\begin{figure}[h]
	\centering\includegraphics[width=10cm]{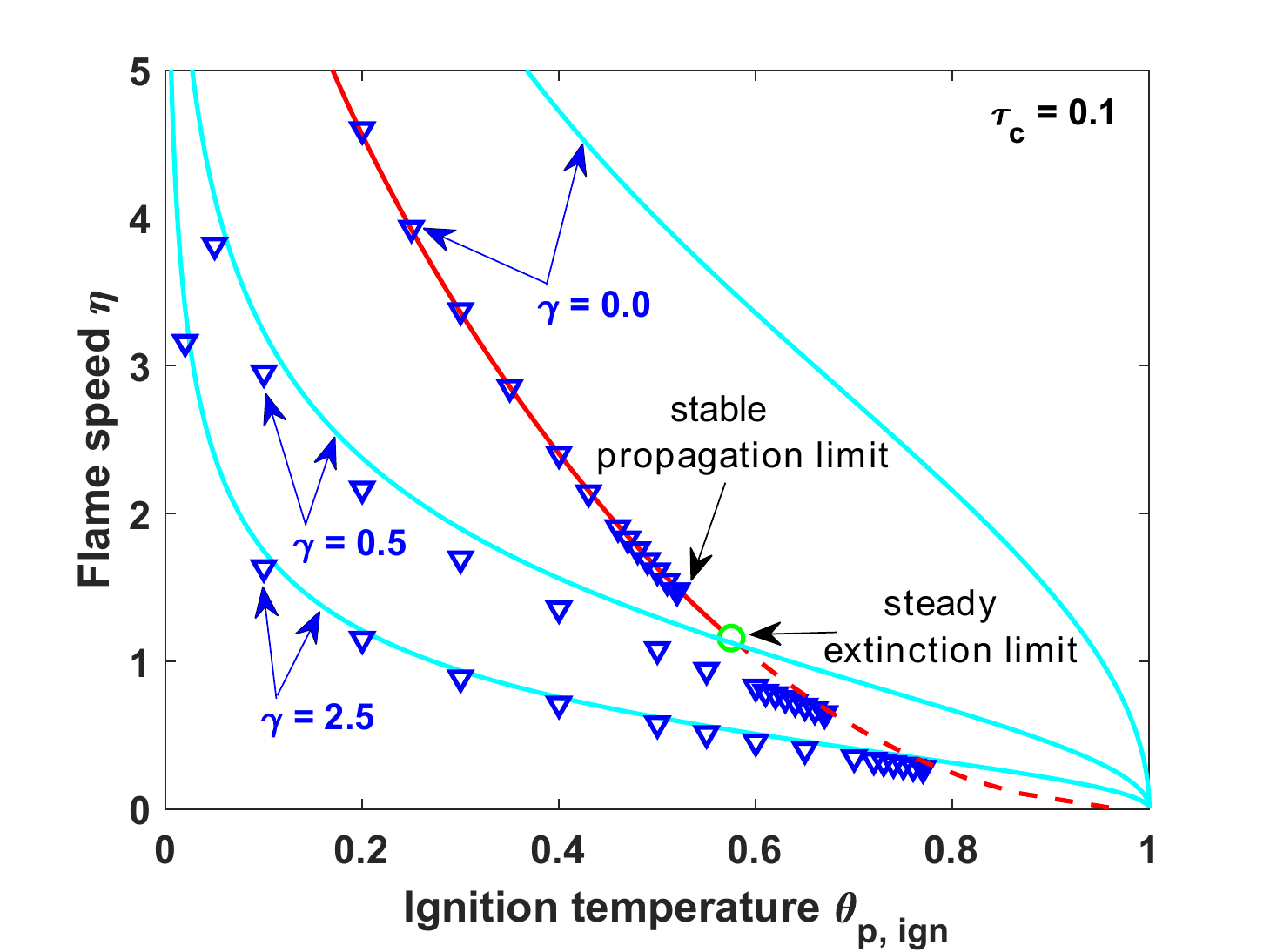}
	\caption{Flame speed as a function of ignition temperature; blue triangles represent transient simulation results, and the dynamical propagation limits at $\gamma=0.0,\ 0.5,\ 2.5$ for $\tau_c=0$ are indicated by solid triangles; red solid and dashed lines are physical and nonphysical solutions of the steady-state discrete model at $\gamma=0$ for $\tau_c=0$, respectively; the cyan solid line is the solution provided by the continuous model; the green open circle is the steady propagation limit.}
	\label{fig:flame speed at tau_c of 0.1}
\end{figure}

\Cref{fig:flame speed at tau_c of 0.1} shows a comparison between the flame speeds obtained by DTIM and CTIM for a representative discrete flame with $\tau_c=0.1$ where particles burn in a finite duration and a large spatial discreteness exists ($\tau_c\ll1$). Selecting a small, but finite combustion time allows a consistent comparison with the prediction of CTIM at different values of  the particle thermal inertia, as at $\tau_c = 0$, CTIM yields infinite values of the flame speed. At $\gamma =0$, the numerical calculation gives practically the same results as the analytical solution of DTEM.  Moreover, CTEM significantly over-predicts the flame speed, which has been already discussed by Shoshin et al. \cite{shoshin1987flame} and Goroshin et al. \cite{goroshin1998effect}. With the increase of particle thermal inertia, the flame speed obtained by numerical simulation using DTIM decreases. At the same time, the numerically computed stable propagation limit extends to a higher ignition temperature approaching the nonphysically analytical solution of DTEM, i.e., the $\gamma=0$ case, which is similar to the scenario where particles burn instantaneously (cf. \cref{fig:flame speed at tau_c of 0}).  Another important finding is that the analytical solution by CTIM and the the numerical solution by DTIM deviate less and less with increasing particle thermal inertia.  These two models almost converge at $\gamma=2.5$ or larger.

To explain the agreement of CTIM and DTIM at large particle thermal inertia, a time scale analysis will be given. First, the failure of the CTEM to correctly predict flame speed at $\tau_c\ll 1$ will be briefly discussed for better understanding.

In DTEM, which neglect the particle thermal inertia, two processes are considered: (i) particle heat release (or combustion) characterized by $t_c$ and (ii) inter-particle heat diffusion characterized by $t_d$ whereas in CTEM only (i) is considered assuming $t_c \gg t_d$ (non-dimensional form: $\tau_c = t_c/t_d \gg 1$). Therefore, CTEM can predict the flame speed that agrees with DTEM when $\tau_c \gg 1$. However, when the spatial discreteness becomes larger ($\tau_c \lesssim 1$), corresponding to a smaller dimensionless burn time $\tau_c$, CTEM over-predicts flame speeds compared to DTEM because now the inter-particle heat diffusion starts to play a role in controlling the flame propagation and will be the dominant process that limits the flame speed when ($\tau_c \ll 1$). As can be seen from \cref{fig:flame speed at tau_c of 0.1}, the prediction of CTIM becomes better with increasing $\gamma$ even at a large spatial discreteness ($\tau_c = 0.1$) although the continuum assumption (i.e., $\tau_c \gg 1$) is still involved. As the physical representation of $\gamma$ is the ratio between $t_e$ (characteristic heating time of a particle in hot gas) and $t_d$ (heat diffusion time in gaseous medium between neighboring particles), one more physical process, i.e., gas-particle heat exchange in the preheating zone, is included in both discrete and continuous models after considering the particle thermal inertia. When $t_e > t_d > t_c$ (normalizing it by $t_d$ gives a dimensionless form: $\gamma > 1 > \tau_c$ ) or $t_e > t_c > t_d$  (dimensionless form: $\gamma > \tau_c > 1$) the flame speed is controlled by the gas-particle heat exchange rate. As in both CTIM and DTIM, particle thermal inertia is included,  they will finally give converged results when the gas-particle heat exchange becomes the limiting process. This result shows that the condition, $\tau_c \gg 1$, becomes unnecessary for the applicability of the continuous model if the particle thermal inertia is sufficiently large.

To demonstrate that CTIM could be conditionally applicable for the prediction of flame speeds without fullfilling $\tau_c \gg 1$ , the temperature histories of a particle and the local gas calculated by DTIM and CTIM will be compared for both small and larger values of the particle thermal inertia. For a steady-state flame, the spatial coordinates  used in CTIM can be converted to time via: $y=x/(v_ut_c)=\tau/\tau_c$. Therefore, the temperature profiles of a particle and the local gas in the preheating zone can be expressed as a function of time:

\begin{equation}
	\theta_p =\theta_{p,ign}\exp(\tau\eta^2), 
\end{equation}
\begin{equation}
	\theta_g =(1+\eta^2\gamma)\theta_p. 
\end{equation}
The particle reaches the ignition temperature at the moment of $\tau = 0$.
In DTIM, the gas temperature $\theta_g$  at the location of a particle as a function of time $\tau$ and a known flame speed $\eta$ can be calculated using the following analytical expression:

\begin{equation}
	\theta_g(\tau, \eta)=
	\begin{cases}
		\displaystyle  \sum_{i= s }^{\infty}\left(\frac{1}{4\pi\tau_r}\right)^{1/2} \exp(-\frac{i^2}{4\tau_r}), & \text{if}\ \tau_c=0 \\
		
		\displaystyle \sum_{i=s}^{\infty} \frac{1}{\tau_c}\int_{(\tau_r-\tau_c)\textbf{\emph{H}}(\tau_r-\tau_c)}^{\tau_r} \left( \frac{1}{4\pi \tau}\right) ^{1/2} \exp(-\frac{i ^2}{4\tau})d\tau, & \text{if}\ \tau_c>0
	\end{cases}	
	\label{eqn: temperature history by DTIM}
\end{equation}
where $s = \textbf{ceil} \left( r-\tau \eta \right)$ and \textbf{ceil} is the ceiling function; $\tau_r = (i-r)/\eta + \tau$ and $\tau$ is the time elapsed since the ignition of the $r$-th previous particle. The temperature history of the particle at the same location can be obtained by numerical integration of \cref{eqn: dimensionless discrete particle temperature equation 1}. Note that the $r$-th previous particle should be far away so that the initial temperature of the current particle is 0 at the ignition moment of the $r$-th particle, which is needed as the initial condition for the numerical integration. It was empirically found that $r=10$ is already sufficient, and there is no noticeable change by further increasing it. In order to compare with the results by CTIM, the ignition moment calculated by \cref{eqn: temperature history by DTIM} is shifted to $\tau=0$.

\begin{figure}[h]
	\centering\includegraphics[width=13.5cm]{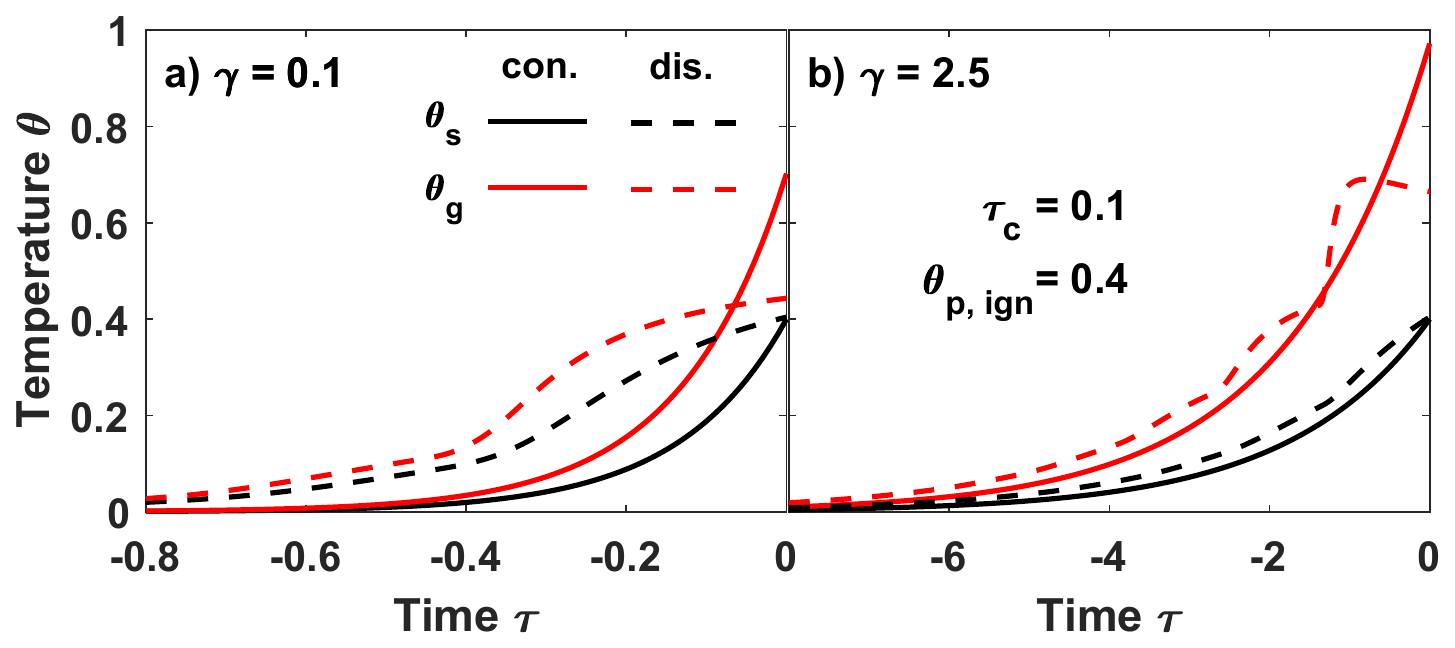}
	\caption{Comparisons between temperature histories of a particle and local gas in the preheating zone predicted by CTIM and DTIM for small (a) $\gamma= 0.1$ and large (b) $\gamma=2.5$ particle thermal inertia. The particle reaches the ignition temperature at $\tau=0.$ The particle combustion time $\tau_c=0.1$, and the particle ignition temperature $\theta_{p,ign}=0.4$.}
	\label{fig:preheating zone structure}
\end{figure}

\Cref{fig:preheating zone structure} shows comparisons between CTIM and DTIM on predicting the temperature evolution of a particle and the local gas in the preheating zone. Here, a small particle burn time, $\tau_c = 0.1$, is chosen deliberately to violate the continuum  assumption. \Cref{fig:preheating zone structure}(a) shows that, at a small value of particle thermal inertia, $\gamma=0.1$,  DTIM predicts that the particle temperature changes similarly to that of the local gas due to the fast response of the particle. The temperature histories predicted by CTIM deviate considerably from those predicted by DTIM.  
\Cref{fig:preheating zone structure}(b) shows that,  at a large value of particle thermal inertia, $\gamma=2.5$, the local gas temperature history, given by CTIM close to the ignition moment, deviates from the one predicted by DTIM. However, due to slow particle response to the changing gas temperature, this deviation is averaged over a period of time, and the particle heating history given by CTIM becomes similar to the one predicted by DTIM. Thus, at large particle thermal inertia, CTIM can be used for the correct prediction of flame speed (but not for temperature of gas near the flame front) even though it violates the assumption that the control volume contains a large number of burning particles, which is, in contrast, a necessity for the validity of CTEM.

\subsection{Propagation properties of thermal inertia controlled flames }

\begin{figure}[h]
	\centering\includegraphics[width=10cm]{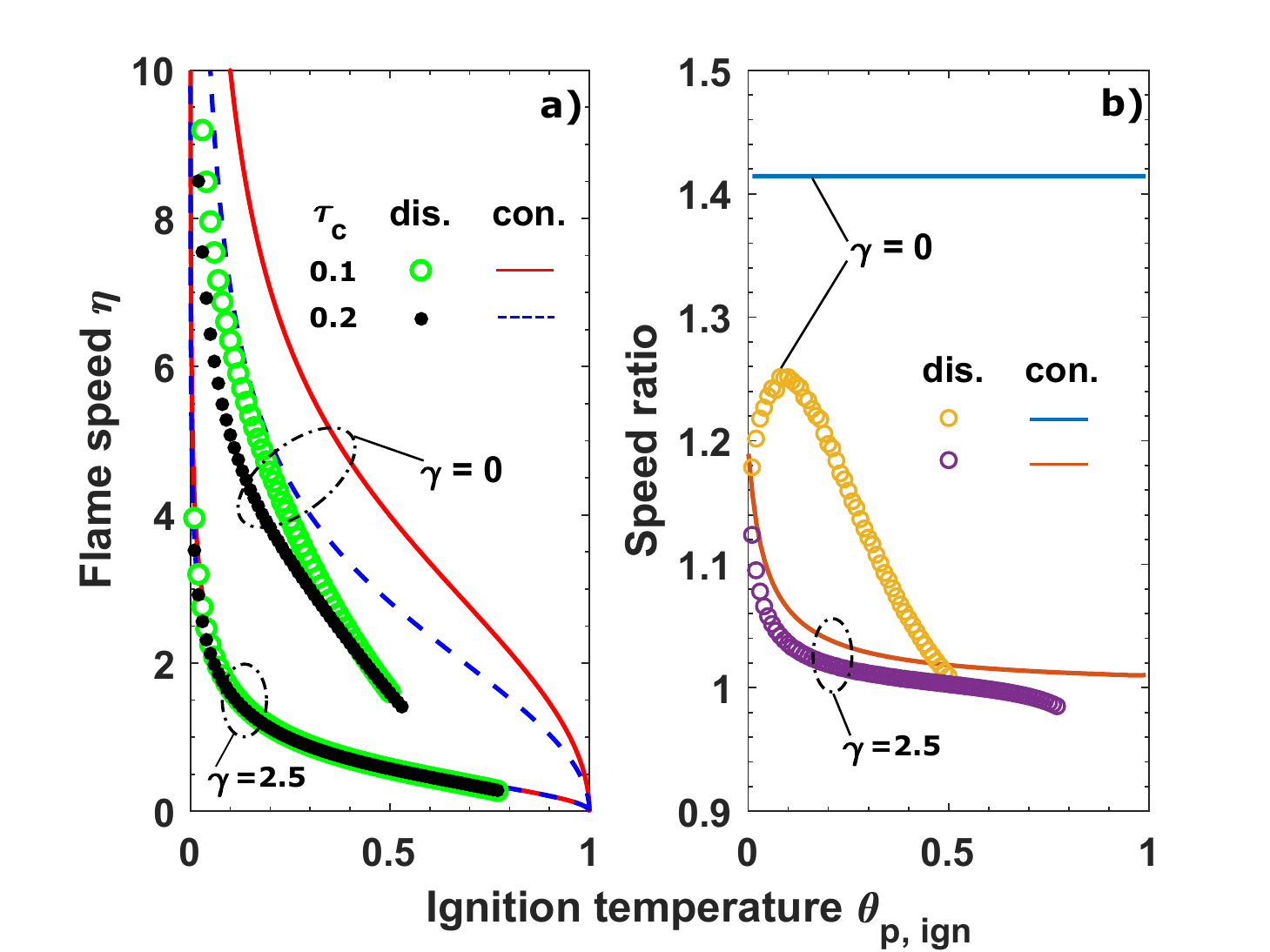}
	\caption{Flame speeds (a) and the their ratios (b) as a function of particle ignition temperature for fast burning particles ($\tau_c = 0.1$  and $\tau_c = 0.2$) calculated using CTIM and DTIM with and without thermal inertia.}
	\label{fig: flame speed ratio 1}
\end{figure}

CTEM predicts that the burning velocity is proportional to the square root of reaction rate ($1/\tau_c$): $\eta\propto\sqrt{\kappa(\theta_{ign})/\tau_c}$, whereas DTEM indicates that the effect of particle combustion rate on the burning velocity becomes weaker with the decrease of dimensionless burn time and disappears for the limiting case of $\tau_c \to 0$. \Cref{fig: flame speed ratio 1}(a) shows that, at $\gamma=0$, CTIM not only over-predicts the burning velocity but also suggests a higher sensitivity of the flame speed (larger flame speed difference at the same ignition temperature) to the particle burn time as compared to DTIM when $\tau_c \ll 1$. It is also seen in \cref{fig: flame speed ratio 1}(a), that doubling particle combustion time has a very weak effect on the flame speed in both CTIM and DTIM at sufficiently large particle thermal inertia. Without including the particle thermal inertia, such effect (though less prominent, as seen in \cref{fig: flame speed ratio 1}(a)) is predicted only by the discrete models (in this case, DTIM is the same as DTEM) at relatively large particle ignition temperatures for fast burning particles.
 
To examine the response of the flame speed to the particle burn time quantitatively, the  ratios between flame speeds at $\tau_c = 0.1$ and 0.2 are calculated using both CTIM and DTIM. \Cref{fig: flame speed ratio 1}(b) shows that, at $\gamma = 0$, CTIM predicts that the flame speed increases $\sqrt{2}$ times when the burn time halves, and the speed ratio is independent on the ignition temperature. At the same time, DTIM suggests a smaller speed ratio, which varies between around 1.0 and 1.26 with the particle ignition temperature and approaches unity when $\theta \approx 0.5$. Here, the relatively weak dependence of the flame speed on the  particle burn time, predicted by DTIM, as compared to CTIM, only results from the discrete nature of heat sources. When  the particle thermal inertia is considered and is sufficiently large, CTIM and DTIM predict very close flame speeds as already mentioned in the previous section. More importantly, the speed ratio calculated by CTIM is not constant. Instead, it also varies between 1.0 and 1.2 with the changing ignition temperature and is slightly larger than but very close to the values predicted by DTIM. After including the particle thermal inertia, the speed ratios calculated using CTIM and DTIM become considerably smaller than their zero-thermal-inertia counterparts and approach unity at a relatively small particle ignition temperature. This means that, besides the spatial discreteness of heat sources, the particle thermal inertia can also make the flame speed less sensitive to the particle burn time. As CTIM is not able to detect the influence of spatial discreteness, the weak response of flame speed to particle burn time at $\gamma=2.5$ is only attributed to the particle thermal inertia. Furthermore, the contribution of the spatial discreteness to the observed weak sensitivity of flame speed is minimal when $\gamma \gg \tau_c$, which can be seen from the small gap between the speed ratios computed with CTIM and DTIM at $\gamma=2.5$.

\begin{figure}[h]
	\centering\includegraphics[width=10cm]{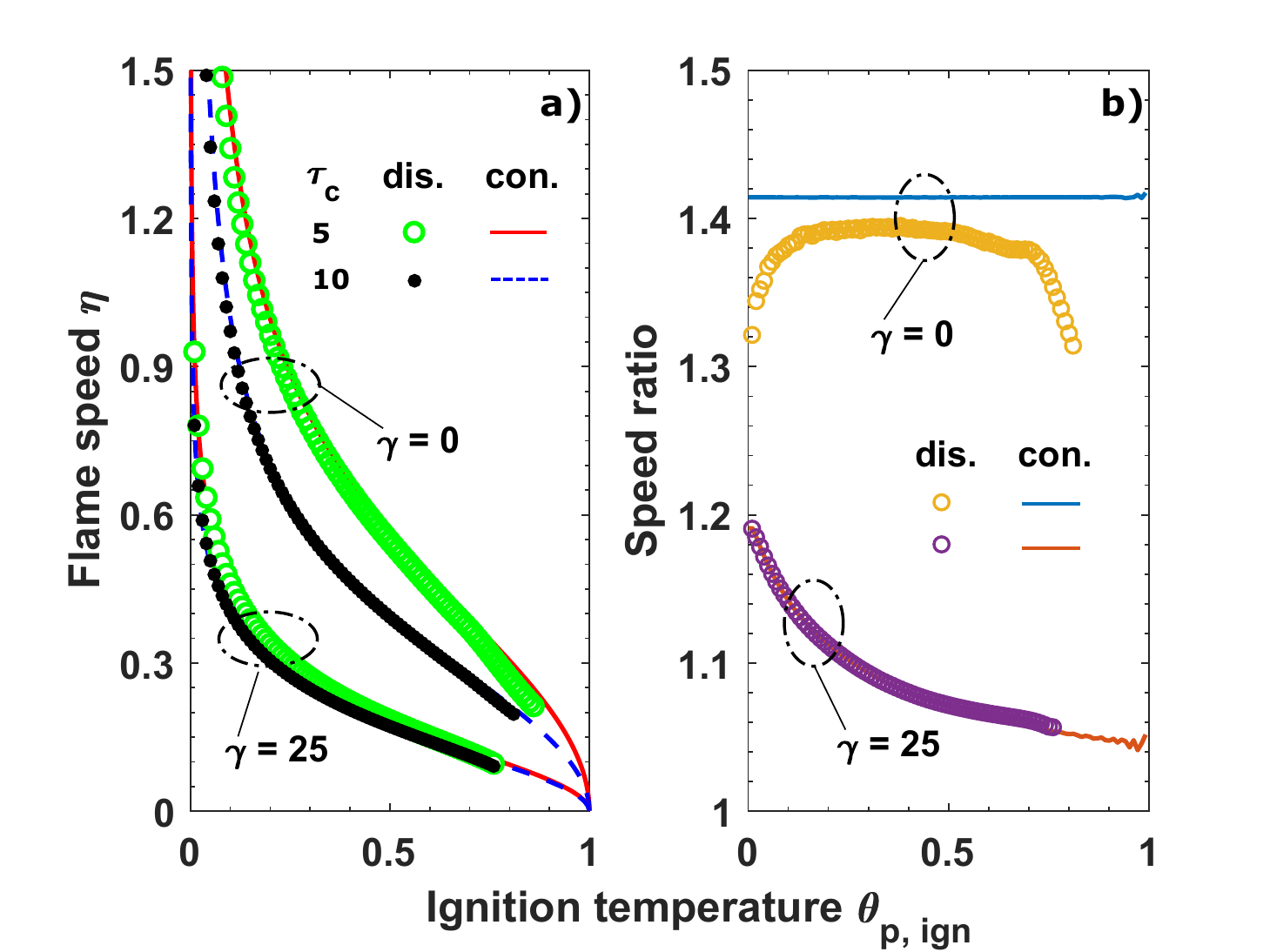}
	\caption{Flame speeds (a) and the their ratios (b) as a function of particle ignition temperature for slowly burning particles ($\tau_c = 5$  and $\tau_c = 10$) calculated using CTIM and DTIM with and without thermal inertia.}
	\label{fig: flame speed ratio 2}
\end{figure}

As already been discussed, when the dimensionless combustion time, which is the only discrete parameter in DTEM, is sufficiently larger than unity,  both DTEM and CTEM give almost the same flame speed because the effect of the spatial discreteness disappears \cite{goroshin1998effect}. In other words, the flame propagates in the continuous regime. \Cref{fig: flame speed ratio 2}(a) shows that regardless of the value of the particle thermal inertia, CTIM and DTIM predict very close flame speeds. 
Furthermore, \cref{fig: flame speed ratio 2}(b) shows that the flame speed ratio calculated using the DTIM at $\gamma = 0$ (equivalent to DTEM) is in close agreement with the value, $\sqrt{2}$, predicted by CTIM at $\gamma  = 0$ (equivalent to CTEM) except at very small and very large particle ignition temperatures. The small difference in the calculated speed ratios between the two models originates from the weak spatial discreteness effect since the selected burn times ($\tau_c=5$ and $\tau_c =10$) do not fully satisfy the required condition of $\tau_c \gg 1$ for the discreteness effect to vanish. Under such conditions, the particle thermal inertia will be the only source making 
the flame speed less sensitive to the particle burn time. To magnify the effect of the particle thermal inertia, a large value of the particle thermal inertia, i.e., $\gamma = 25$ is selected. Similar to a discrete combustion wave, the propagation velocity of a continuous combustion wave becomes less sensitive to the particle burn time as indicated by the computed speed ratio when the particle thermal inertia limits the flame propagation, and the flame propagation velocity will be independent on the reaction rate when $\gamma \gg \tau_c $. 


The abovementioned results suggest that the weak dependence of flame speed on particle burn time can be attributed to the spatial discreteness of heat sources or the thermal inertia of particles or both of them in the case of short particle combustion time. Often, the dependence of flame speed on particle burn time, evaluated by a flame speed ratio, is used to experimentally identify the flame propagation regime for particular suspensions (discrete v.s. continuous regime) \cite{wright2016discrete, palevcka2019new}. However, when the particle thermal inertia is a controlling parameter for the flame speed, both of the predicted burning velocities and speed ratios by CTIM and DTIM are in a good agreement. Moreover, in the case of long combustion times, large particle thermal inertia leads to not only better agreement between flame speeds calculated with CTIM and DTIM, but also to a very weak sensitivity of flame speed to particle combustion time. Therefore, the experimentally observed weak dependency of flame speed on particle burn time is not sufficient to prove the occurrence of discrete flames without special attention paid to the particle thermal inertia when interpreting experimental results of flame propagation speeds. 

\subsection{Flame dynamic regimes and detailed transient propagation behavior}
\label{subsec 3.3}
Besides the influence of the particle thermal inertia on stable flames, it is also found that the particle thermal inertia changes the dynamical behavior of flames consisting of fast burning particles ($\tau_c \ll 1$). In this section, sample results of flame dynamic regimes and transient propagation behavior will be presented in detail for instantly burning particles.

\begin{figure}[h]
	\centering\includegraphics[width=13.5cm]{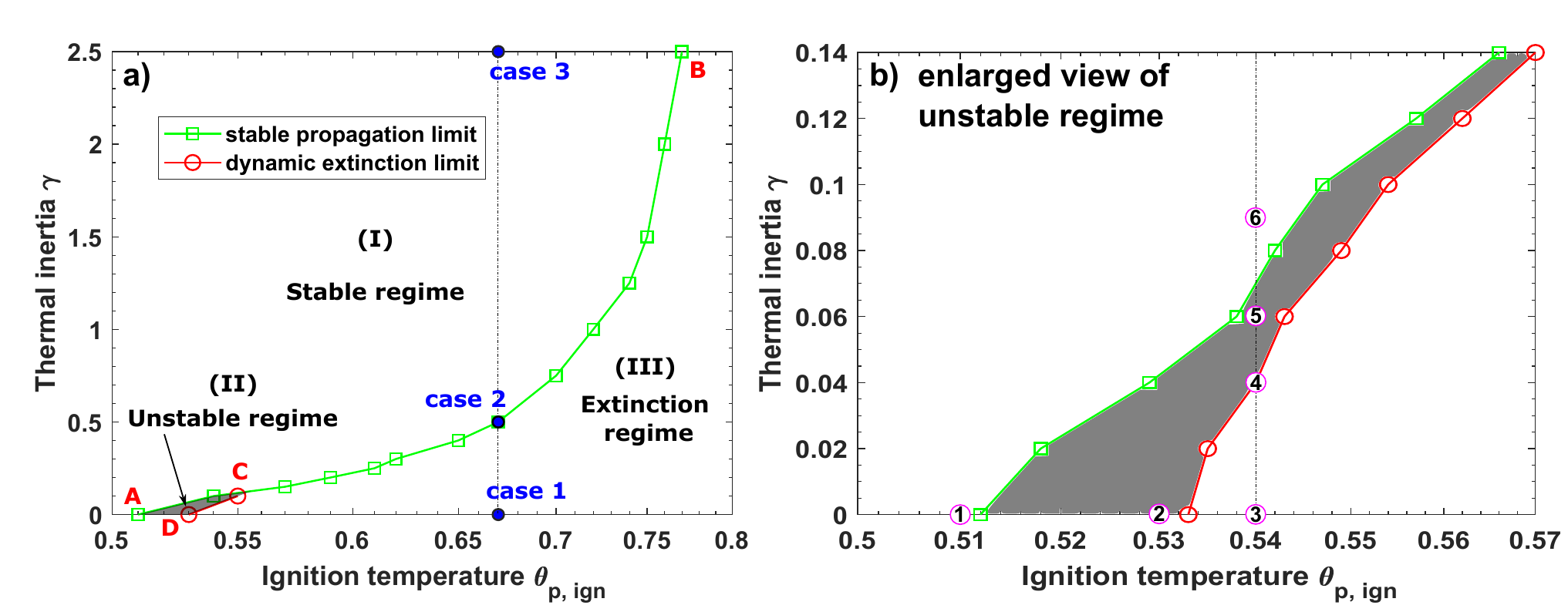}
	\caption{Regime diagram of flame dynamics (a) and enlarged view of unstable regime (b) for one-dimensional flames consisting of instantly burning particles ($\tau_c =0$).}
	\label{fig:limit digram}
\end{figure}

\Cref{fig:limit digram}(a) shows the diagram of flame dynamics for flames propagating in a one-dimensional regular array of instantly burning particles ($\tau_c = 0$). The limits (critical ignition temperature) are obtained by gradually increasing the ignition temperature with the step of 0.01 in the simulations using DTIM. The stable propagation limit increases from 0.51 to 0.77 when particle thermal inertia increases from 0 to 2.5. On the left side of the boundary line AB, all flames propagate in the stable regime. At small particle thermal inertia, $0 \leqslant \gamma \leqslant 0.14$, flames becomes unstable and are still able to propagate through the whole domain when the ignition temperature is beyond the stable propagation limit (AB), until reaching the dynamic extinction limit (CD). The shaded region ACD corresponds to the unstable regime. \Cref{fig:limit digram}(b) shows a refined result and an enlarged view of the unstable regime. A smaller ignition temperature step of 0.001 is used for the refined boundaries between flame dynamic regimes in the enlarged view. As shown in \cref{fig:limit digram}(b), with the increasing particle thermal inertia, the unstable regime narrows. Observed that at $\gamma$ values larger than 0.14, unstably propagating flames also exist at ignition temperatures slightly exceeding the stable propagation limit, but the width of the unstable region is very narrow ($<0.005$). Therefore, it is neglected in \cref{fig:limit digram}(a). When the particle ignition temperature is beyond the boundary BCD, flames propagate in the extinction regime because they quench before reaching the end of the computational domain.

\if
Alternatively, the identical temperature histories of a particle and the corresponding local gas can also be obtained from the numerical simulation for the stable propagating flames. The gas ignition temperature obtained from post-processing and simulation are compared in Table 1, which shows a negligible discrepancy.
\fi

\begin{figure}[h]
	\centering\includegraphics[width=13.5cm]{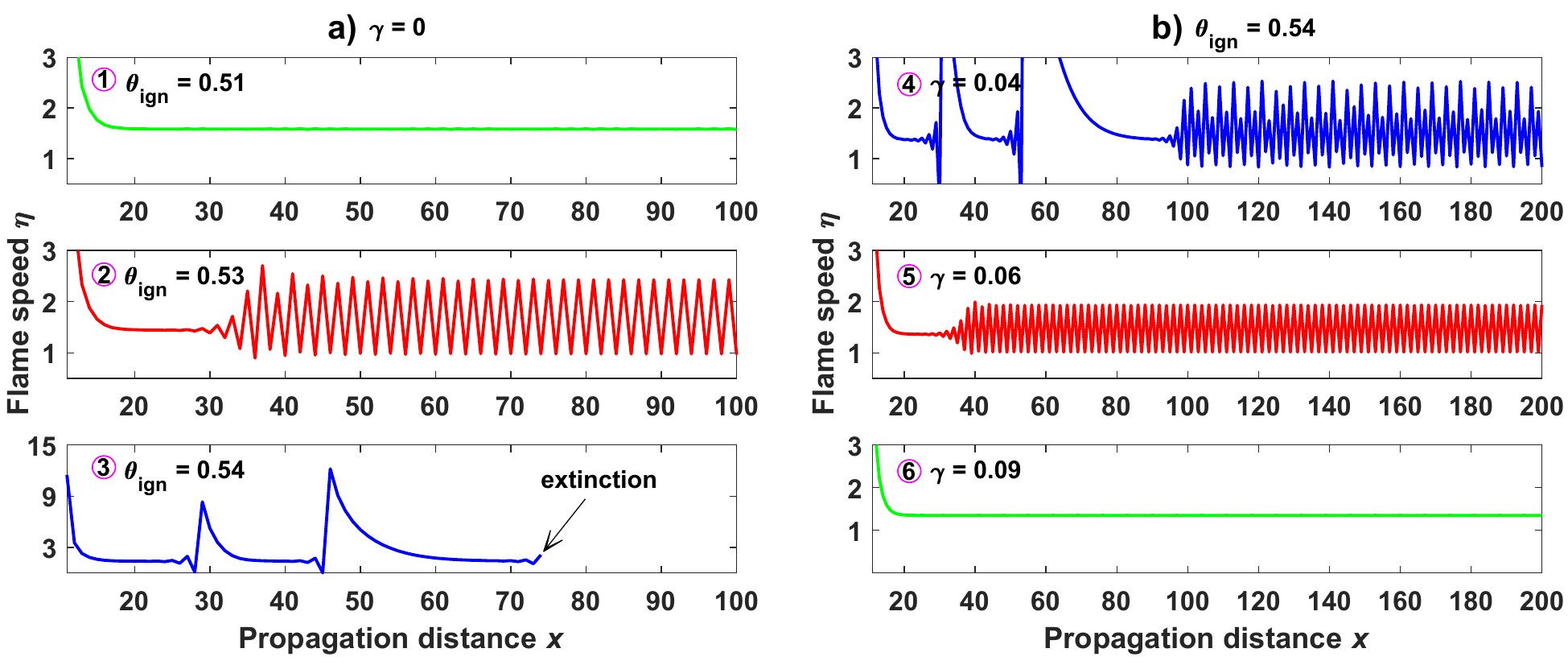}
	\caption{Flame speed as a function of propagation distance for cases without particle thermal inertia (a) at three different particle ignition temperatures of \textcircled{1} $\theta_{ign}=0.51$, \textcircled{2} $\theta_{ign}=0.53$ and \textcircled{3} $\theta_{ign}=0.54$,  and for cases with the particle thermal inertia values (b) of \textcircled{4} $\gamma=0.51$, \textcircled{5} $\gamma=0.06$, and \textcircled{6} $\gamma =0.09$  at a fixed particle ignition temperature of $\theta_{ign}=0.54$. The flame dynamic regimes of these cases are marked in Fig.6(b).}
	\label{fig: flame dynamics withou thermal interia}
\end{figure}
When the particle ignition temperature exceeds the stable propagation limit, complex flame dynamics is observed for both cases of $\tau_c=0$ and $\tau_c>0$ without considering the particle thermal inertia \cite{tang2009effect, rashkovskiy2010one}. To demonstrate the flame dynamics, sample results are selected at $\tau_c=0$ and $\gamma = 0$, as marked by \textcircled{1}, \textcircled{2}, and \textcircled{3} in \cref{fig:limit digram}(b). \Cref{fig: flame dynamics withou thermal interia}(a) shows that at $\theta_{ign}=0.51$,  the flame becomes stable after a short propagation distance that is influenced by the numerical initiation for the flame. At a slightly larger ignition temperature, $\theta_{ign}=0.53$, the flame first behaves similarly to the flame of $\theta_{ign}=0.51$ and a short stage of a stable flame between $\bold {x} = 20$ and 30 exists, thereafter the flame becomes intrinsically unstable with a varying fluctuation amplitude in flame speed between approximately $\bold{x} = 30$ and 50. Finally, a flame with a periodically oscillating speed is observed. In this regime, a flame can propagate through the whole computational domain (by further increasing the domain size, the manner of the flame propagation does not change). When the ignition temperature is further increased to $\theta_{ign}=0.54$, the flame extinguishes at approximately $\bold x = 75 $ after two large fluctuation.  
Applying the particle thermal inertia changes not only the dynamic propagation regime of a flame, but also detailed flame behavior such as periodical oscillations and chaotic fluctuations, as will be illustrated by sample cases marked by \textcircled{4}, \textcircled{5}, and \textcircled{6} in \cref{fig:limit digram}(b). \Cref{fig: flame dynamics withou thermal interia}(b) shows that, when the small particle thermal inertia of 0.04 is applied to particles, the flame first undergoes two round large fluctuations between $\bold x = 20$ and 90. Hereafter the flame propagates with chaotic fluctuations of the flame speed over the rest of the domain. However, without considering the particle thermal inertia, the flame extinguishes at the same ignition temperature $\theta_{ign}=0.54$ as illustrated in \cref{fig: flame dynamics withou thermal interia}(a)\textcircled{3}. With further increasing the particle thermal inertia to $\gamma = 0.06$, the flame becomes able to propagate through the entire domain in a manner of periodical fluctuations. With the increasing particle thermal inertia to 0.09, the flame becomes stable.   
Therefore, for flames propagating in a one-dimensional array of regularly spaced fast burning particles, the particle thermal inertia is an important fact that needs to be considered to predict the propagation limits and dynamical flame behavior.

\subsection{Mechanism of extended propagation limits}
As already shown in \cref{fig:flame speed at tau_c of 0} for instantly burning particles and \cref{fig:flame speed at tau_c of 0.1} for particles with finite combustion rate, the particle thermal inertia extends the stable propagation limits although the flame speed decreases. In order to understand the mechanism of this phenomenon, 
three cases with the same particle ignition temperature of 0.67 and different particle thermal inertia values of 0, 0.5, and 2.5, respectively,  are selected to examine the temperature evolution of a particle and local gas. The temperature histories will manifest the mechanism of the extended stable propagation limits by particle thermal inertia. The 3 cases belong to 3 different flame dynamic regimes as marked in \cref{fig:limit digram}(a).

\begin{figure}[h!]
	\centering\includegraphics[width=7.5cm]{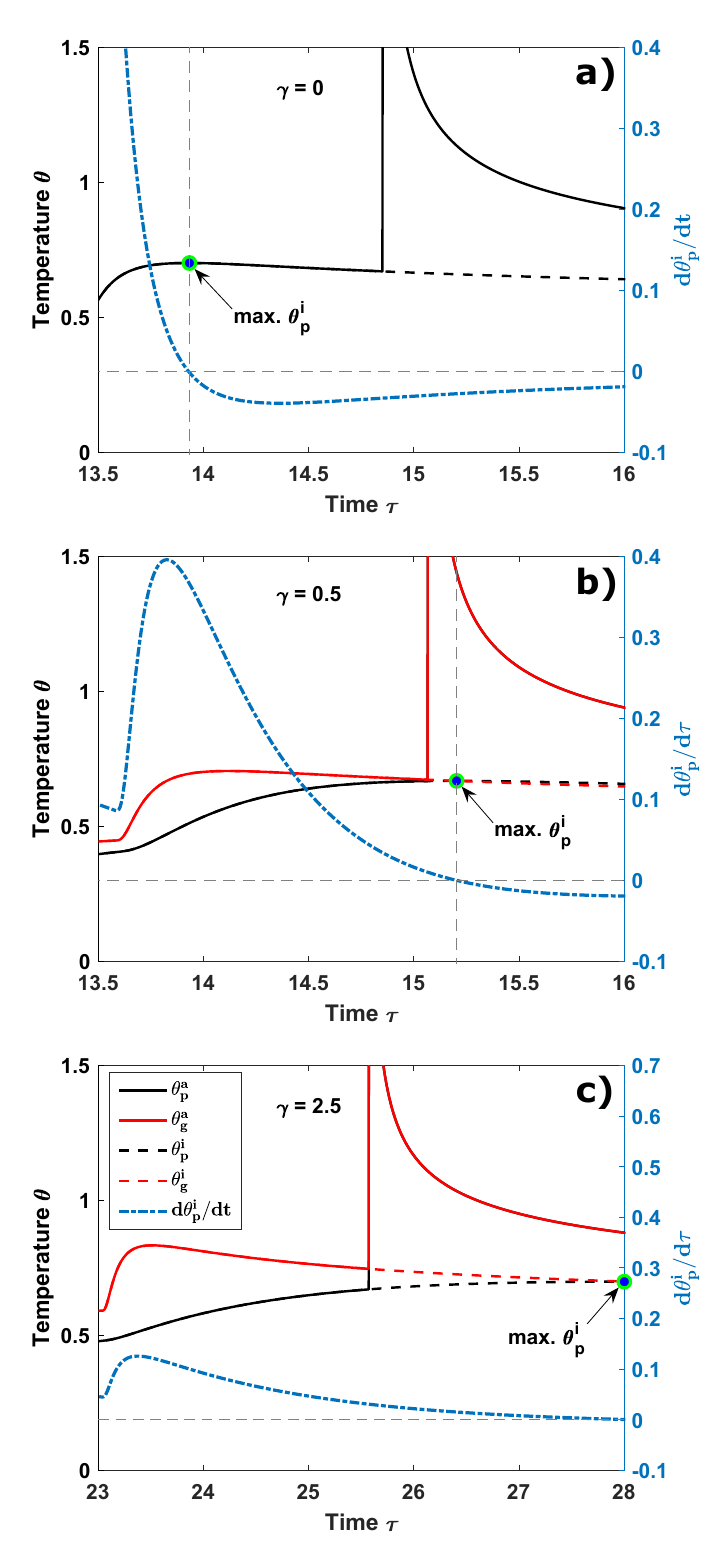}
	\caption{Temperature histories of a reactive particle $\theta_p^a$ (solid black line) and corresponding gas $\theta_p^a$ (solid red line), and temperature histories of an inert particle $\theta_p^i$ (dashed black line) and corresponding gas $\theta_p^a$ (dashed red line) in the case of a) $\gamma = 0$, b) $\gamma = 0.5$, and c) $\gamma = 2.5$ and   $\tau_c =0$.Particle ignition temperature $\theta_{p,ign}=0.67$}
	\label{fig: Temperature history}
\end{figure}

Case 1 belongs to the extinction regime where no stable flame propagation velocity can be obtained from transient numerical simulations but an analytical solution (nonphysical) of flame speed for this condition ($\gamma = 0$) still exists when the assumption of steady-state flame is adopted. The analytical solution is substituted into \cref{eqn: temperature history by DTIM} to obtain the temperature evolution of both a particle and local gas. 
This nonphysical solution does not mean that the flame can actually propagate, and when propagation is not possible the analytical flame speed means that reacting and reacted particles are ``forced" to ignite at a periodic time interval. Although the assumption of steady-state flame may not be physical at some parameters, it allows for better understanding of the mechanism by which the flame propagation limits are extended when the particle thermal inertia is considered. The heat release of the current particle can be numerically suppressed by dropping the term of $i=0$ in \cref{eqn: temperature history by DTIM}, which results in the temperature history of the local gas with an inert particle placed at the same position. Then, the temperature evolution of the inert particle can be obtained in the same manner as for reactive  particles. \Cref{fig: Temperature history} (a) shows the temperature histories of a reactive ($\theta_p^a$) and an inert ($\theta_p^i$) particles for the case of $\gamma=0$. The first order derivative  of $\theta_p^i$ with respect to time $\tau$ is calculated to locate the maximum of $\theta_p^i$ ($\theta_{p,max}^i$), which is
also shown in the same figure. As can be seen, the particle passes its ignition temperature (indicated by the horizontal red dash-dot line) at the first time without ignition. After $\theta_{p,max}^i$, the reactive particle ignites when reaching the ignition temperature at the second time during cooling down. In reality, this process is impossible. Therefore, the analytical solution for the flame speed in this case is nonphysical. Furthermore, it is easy to understand that physical ignition is impossible if the ignition temperature is larger than $\theta_{p,max}^i$. Thus, the propagation limit (steady extinction limit) is reached when the particle ignites at $\theta_{p,max}^i$ if without considering any dynamic effects. \Cref{fig: Temperature history} (b) shows the temperature histories of a reactive particle and an inert particle as well as that of the corresponding local gas for $\gamma = 0.5$ and $\tau_c = 0$. This case is located  exactly at the stable propagation limit as shown in \cref{fig:limit digram}. Due to the thermal inertia, the particle can still be heated up during cooling of gas until $\theta_g=\theta_p$, when the inert particle reaches $\theta_{p,max}^i$. As can be seen, the reactive particle takes more time (around 15.1) to reach the ignition temperature compared to the case 1, and this happens slightly before reaching $\theta_{p,max}^i$. In transient simulations,  it is impossible that particle ignition occurs exactly at $\theta_{p,max}^i$ (i.e. steady extinction limit) because a flame quenches due to the dynamic effect (i.e. unbounded flame speed fluctuations), as discussed in \cref{subsec 3.3}, when the particle ignition temperature is slightly smaller than $\theta_{p,max}^i$. When thermal inertia $\gamma$ is further increased to 2.5 (case 3), the reactive particle ignites much earlier before $\theta_{p,max}^i$ and the gas ignition temperature is obviously higher than the particle's as shown in \cref{fig: Temperature history} (c). Therefore, DTEM underestimates flame propagation limits because no physical ignition can be predicted once gas temperature starts to decrease, while in practice, owing to the particle thermal inertia, particle responds slowly to gas temperature and can still be heated up to ignite during cooling of gas.

As already shown in \cref{fig:flame speed at tau_c of 0,fig:flame speed at tau_c of 0.1}, the stable propagation limits of one-dimensional discrete combustion wave predicted using DTIM always approach the nonphysical solution of flame propagation speed found without considering the thermal inertia. As discussed above, the flame reaches its steady-state propagation limits when the particle ignition temperature is equal to $\theta_{p,max}^i$. By substituting $\frac{d\theta_{p}}{d\tau}=0$ into \cref{eqn: dimensionless discrete particle temperature equation 1}, we see that the particle reaches $\theta_{p,max}^i$ when the local temperatures of the gas and the particle are equal: $T_p = T_g$. Therefore, if the particle ignites at this moment, the particle ignition temperature is identical to the gas ignition temperature.  The comparison between cases 2 and 3 in \cref{fig: Temperature history} (b) and (c) clearly shows that the ignition temperatures of particle and gas become closer when the flame approaches to the stable propagation limits (see also Fig. 6(a)). Therefore, at the propagation limits, particle and local gas reach thermal equilibrium and the gas ignition temperature used in DTEM can be considered as the particle ignition temperature. Thus, the nonphysical solution branch of the traditional discrete model is actually a set of solutions for the propagation limits (steady extinction limits) of steady-state combustion waves with different thermal inertia. The steady extinction limits deviate only minimally from the 
stable propagation limits calculated numerically using the DTIM, which can also be seen from \cref{fig:flame speed at tau_c of 0,fig:flame speed at tau_c of 0.1}. 

\subsection{Application: new interpretation of the micro-gravity experiment}

In this section, the model developed in this paper, i.e., DTIM, will be applied to predict the flame propagation speed in the suspensions of iron particles, for which experiments have been conducted at micro-gravity. Besides, the limitations of applying DTIM will be discussed.

\begin{figure}[h]
	\centering\includegraphics[width=8.5cm]{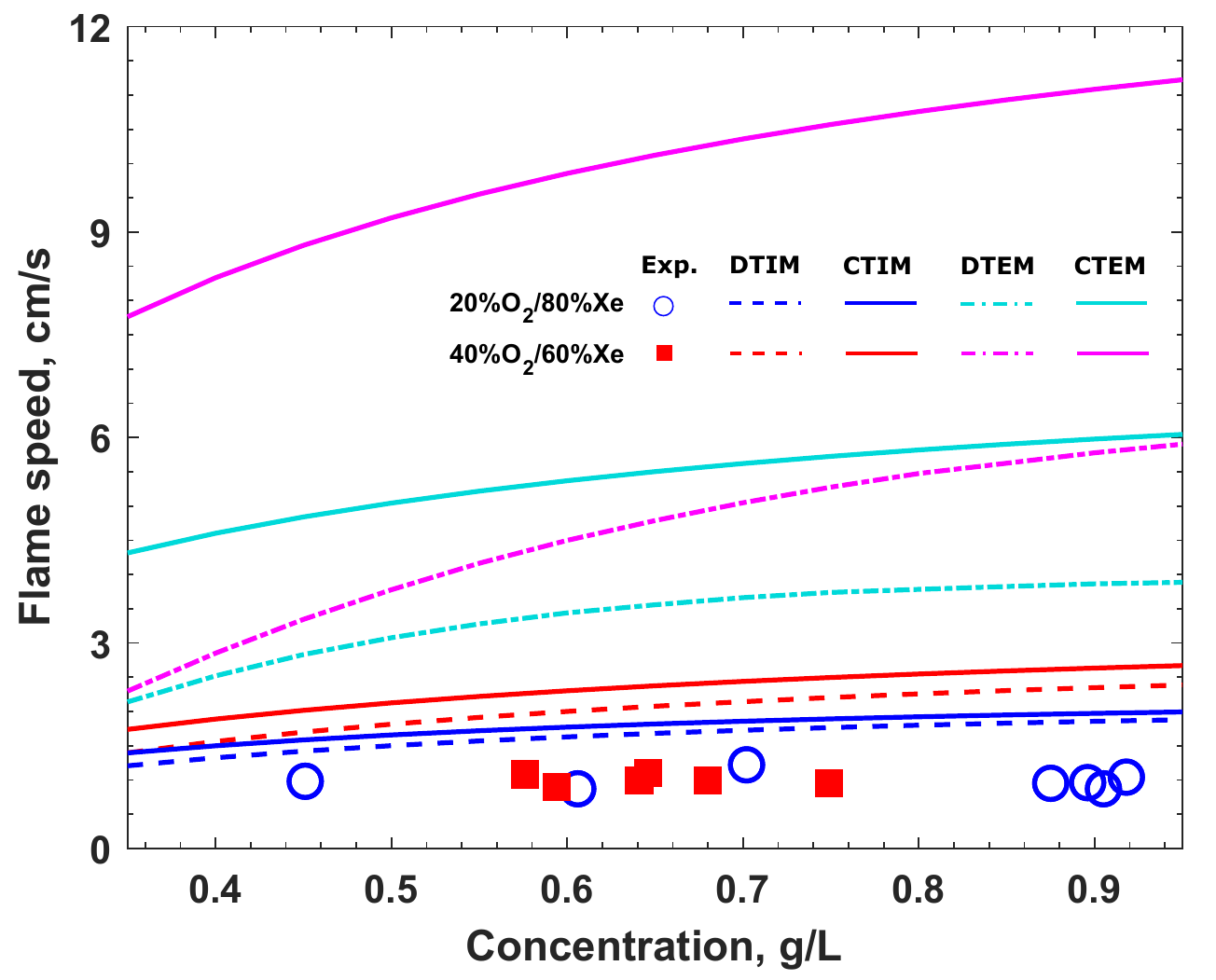}
	\caption{Comparisons between measured flame propagation speeds in suspensions of iron particles ($d_{32} = 33\, \mu$m) in 20\%O$_2$/80\%Xe and 40\%O$_2$/60\%Xe gas mixtures at micro-gravity \cite{palevcka2019new} and predicted flame speeds by DTEM, CTEM, DTIM, and CTIM, using  parameters corresponding to experimental conditions, as a function of iron particle concentration.}
	\label{fig: application for iron flame}
\end{figure}

Based on the experimental parameters including the thermophysical properties of the gaseous mixtures (20\%O$_2$/10\%Xe and 40\%O$_2$/60\%Xe), particle ignition temperature, and particle burn time, reported in \cite{palevcka2019new} and the bulk properties of iron, dimensionless burn time, $\tau_c$, and dimensionless particle thermal inertia, $\gamma$, are estimated. \Cref{tab: table 1} shows that at the iron-particle concentration of \SI{0.7}{g/L} (fuel lean condition), the estimated values of $\gamma$ are about 6 (at 20\% O$_2$) to 12 times (at 40\% O$_2$) larger than those of $\tau_c$, and the order of magnitude of $\gamma$ is unity. This means that particle combustion occurs much faster than both inter-particle heat diffusion and gas-particle heat exchange and that the later two processes are comparably slow. Thus, the estimated dimensionless parameters preliminarily suggest that the flame propagation is limited by the combined effect of the spatial discreteness and particle thermal inertia. In order to evaluate the importance of the particle thermal inertia to the flame propagation at the experimental conditions reported in \cite{palevcka2019new}, the flame speeds calculated using DTIM and DTEM are compared. 
\Cref{fig: application for iron flame} shows that DTEM overestimates the flame speed 
by more than two-folds compared to the measurements. After considering particle thermal inertia, the prediction made by DTIM is considerably improved not only for the flame speed but also for the the weak dependence of the flame speed on oxygen concentration (or particle burn time). To evaluate the respective contributions of the spacial discreteness and particle thermal inertia to limiting the flame propagation, the flame speed is also calculated using CTIM and CTEM and compared with the prediction of DTIM. As can be seen in \cref{fig: application for iron flame}, the prediction of CTIM is slightly (e.g. about 8\% at \SI{0.7}{g/L} for 20\% O$_2$) larger than that of DCIM due to the uncaptured spacial discreteness effect in CTIM. On the contrary, the prediction of CTIM significantly (e.g. 2.6 times at \SI{0.7}{g/L} for 20\% O$_2$) smaller than the overestimated results of CTEM that ``forgets'' both effects of the spacial discreteness and particle thermal inertia. The quantitative comparison clearly indicates that at the experimental conditions \cite{palevcka2019new}, the flame propagation is mainly limited by particle thermal inertia and therefore, it is insensitive to the particle burn time.

Although in the reaction zone, the particle inertia is not considered in DTIM, it can still give a quantitatively reasonable prediction for the experiment. This is because of the fact that particle thermal inertia in the reaction zone slows down heat release to the gas from burning particles, which can be considered as a slower particle combustion rate (longer effective burn time), and at the experimental conditions, the flame propagation is insensitive to particle burn time. Therefore, DTIM  is able to give a good quantitative prediction except that the flame propagation is limited by particle combustion rate, when the model will overestimate the flame speed.

\begin{table}[ht]
	\caption{Estimation of the values of the dimensionless burn time, $\tau_c$, and dimensionless particle thermal inertia, $\gamma$, for suspensions of iron particles ($d_{32} = 33\, \mu m$) in O$_{2}$/Xe mixtures at the fuel concentration of $B_u=0.7\,$g/L. The gas thermophysical properties, particle burn time and particle ignition temperature (\SI{900}{K}) are adopted from the experimental paper \cite{palevcka2019new}. The specific heat, density and heat of combustion of iron, $c_s = 0 .45\,$J/g$ \cdot$K, $\rho_s = 7874$\,kg/m$^3$, and $Q = 4857\,$J/g are used.} 
	\begin{adjustbox}{width=1\textwidth}
		\begin{tabular}{cccccccccccc}
			\hline
			Gas mixture &$\rho_g$, g/L & $c_g$, J/g$\cdot$K & $\alpha$, cm$^2$/s & $l$, cm & $t_c$, ms & $\theta_{p, ign}$ & $\tau_c$ & $\gamma$ \\ 	\hline
			
			20\%O$_2$/80\%Xe   & 4.6   & 0.21  & 0.059   & 0.067 &   8.1   &   0.23 &   0.13   & 0.71 \\ 
			40\%O$_2$/60\%Xe   & 3.8   & 0.28  & 0.072    & 0.067 &   2.7  &   0.24 & 0.055   & 0.66                  \\ \hline                         
		\end{tabular}
	\end{adjustbox}
	\label{tab: table 1}
\end{table}

\section{Conclusions and outlook}
\label{sec 5}
The current work investigates the thermal inertia effect of point-like condensed reactive particles on the propagation properties, including flame speeds, propagation limits, and near-limit dynamics, of one-dimensional discrete combustion waves using the the extended discrete flame model. The assumption of thermal equilibrium between particles and gas adopted in previous DTEM is removed by introducing a temperature equation for a particle, where the thermal inertia is included as an adjustable parameter.  

After considering particle thermal inertia, a particle in the preheating zone needs more time to reach the ignition temperature due to a smaller  heating rate, which leads to a slower flame propagation velocity. For the discrete combustion waves, propagating in a chain of evenly distributed particles that have a finite combustion rate, the flame speeds predicted by CTIM and DTIM become closer and closer together with the increase of particle thermal inertia, and finally two models converge when the gas-particle heat exchange rate becomes the slowest process limiting the flame propagation. Furthermore, the particle thermal inertia extends the stable propagation limits of the discrete combustion wave because a particle can still be heated up and ignite physically during cooling of local gas as long as gas stays hotter than the particle. With the steady-state assumption, the flame propagation limit (steady extinction limit) is reached when a particle ignites at the maximum temperature that it can possibly obtain during inert heating in the preheating zone. This requirement is meet when $T_{ign,s} = T_{ign,g}$. Therefore, the nonphysical branch of analytical solution for the propagation velocity of the steady-state discrete combustion wave, solved with thermal equilibrium assumption, is a set of solutions for the steady extinction limit of the discrete flames with different thermal inertia. 

The model improved in this paper, i.e., DTIM, also suggests theoretically that the insensitivity of the flame speed to the particle combustion rate can be attributed to particle thermal inertia. Therefore, it may give a misleading result to identify the discrete combustion wave by only evaluating the ratio between flame speeds at different O$_2$ concentrations without considering particle thermal inertia in experiments. When the gas-particle heat exchange is the slowest process, i.e. $\gamma \gg max\left\lbrace 1, \tau_c\right\rbrace$, the flame propagation is limited by the particle thermal inertia instead of the particle burning rate (continuous flame) or the inter-particle heat diffusion (discrete flame). This could be regarded as a new kind of combustion wave: inertia flame.

In practical fuel particle suspensions, particles are located randomly in space and usually polydisperse. Furthermore, particle ignition could be conjugated with chemical kinetics and depend on particle sizes. The combined effects of these features and particle thermal inertia on the propagation speeds and limits of heterogeneous dust flames with relatively large spacial discreteness need to be further explored.
This can be accomplished in future by adopting case-dependent parameters and implementing fuel-specific kinetics, when augmenting the presently developed model to two and three dimensions.

\section*{Acknowledgment}

The authors thank Dr. Xiaocheng Mi for the discussion and suggestions during preparation of the paper.

\appendix

\section{Derivation of the dimensionless form of the governing equation for transient particle temperature}
\label{subsection: appen}
The dimensional governing equation for transient particle temperature is described as:
\begin{equation}
	c_{s}\rho_{s}V_{p}\frac{dT_{p}}{dt}=hA_{p}(T_{g}-T_{p}),
	\label{eqn: discrete particle equation}
\end{equation}
where $T_{p}$ and $T_{g}$ are particle and gas temperature, respectively; $t$ is phycial time; $h=\frac{Nu\lambda_{g,u}}{d_{p}}$ is the heat transfer coefficient, and $Nu$ is the Nusselt number, $d_{p}$ is the particle diameter, $\lambda_{g,u}$ is the gas conductivity; $A_{p}$ and $V_{p}$ are particle surface area and volume, respectively. Therefore, the coefficient $\frac{hA_{p}}{c_{s}\rho_{s}V_{p}}$ has a physical meaning of heating rate, and the inverse, $\frac{c_{s}\rho_{s}V_{p}}{hA_{p}}$ , means the characteristic time scale of the heat exchange between the gas and a particle, $t_{e}$. Since the smaller is the value of $t_{e}$, the faster is the gas-particle  heat exchange, $t_{e}$ can also be interpreted as particle thermal inertia. Using the same dimensionless form for temperature as used in the continuous model, the dimensionless temperatures of the gas and the particle are written as $\theta_{p}=\frac{T_{p}-T_{0}}{T_{a}-T_{0}}$ and $\theta_{g}=\frac{T_{g}-T_{0}}{T_{a}-T_{0}}$, respectively. Besides, we define a dimensionless particle thermal inertia, $\gamma=\left( \frac{\overline{c\rho}V_{p}}{hA_{p}}\right) / \left( \frac{l^{2}}{\alpha_u} \right)=t_{e}/t_{d}$, which clearly has a phycial meaning of the ratio between the characteristic time scales of gas-particles heat exchange and heat diffusion between neighboring particles in the gaseous phase. By assuming Nusselt number $Nu=2$, $\gamma$ can also be written as: 

\begin{equation}
	\gamma=\frac{c_{s}\rho_{s}r_{p}^{2}}{3\overline{c\rho}l^{2}}=\frac{1}{3}\left( \frac{3}{4\pi}\right)^{2/3}\frac{c_s\rho_{s}^{1/3}B^{2/3}}{\overline{c\rho}},
	\label{eqn: dimensionless thermal inertia}
\end{equation}
which is a function of the heat capacities and densities of the gas and the particle as well as the dust concentration. Moreover, \cref{eqn: dimensionless thermal inertia} suggests that $\gamma$ is independent on particle sizes when the same concentration of particles, $B$, is introduced in the system. Finally, \textcolor{black}{by substituting $\theta_{p}$, $\theta_{g}$, and, $\gamma$ into \cref{eqn: discrete particle equation},} the non-dimensional form of \cref{eqn: discrete particle equation} can be expressed as:

\begin{equation}
	\label{non-dimentinal governing euqation for particle}
	\frac{d\theta_p}{d\tau} = \frac{\theta_g - \theta_p}{\gamma},
\end{equation}

\if
While for two-dimensional simulation, 
$\eta_i = (\overline{x}_i-\overline{x}_{i-1})/(\overline{\tau}_{ign,i}-\overline{\tau}_{ign,i-1})$, where $\overline{x}_i$ is the average $x$ coordinate of particles with $x \in [i-0.5, i+0.5)$ and $\overline{\tau}_{ign,i}$ is the average ignition time of these particles. Besides, flame propagates along the $x$ direction. 
\fi

\bibliographystyle{elsarticle-num} 
\bibliography{references}




\end{document}